\documentclass[12pt]{article}
\usepackage{graphics,graphicx}
\usepackage{amssymb,epsfig,amsmath,euscript,array}
\usepackage{cite}
\usepackage{pstricks}
\usepackage{color}
\makeatletter \@addtoreset{equation}{section} \makeatother

\def \shuffle{{\,\amalg\hskip -4.2pt\amalg\,}}
\newcommand \vev [1] {\langle{#1}\rangle}

\newcounter{multieqs}

\newcommand{\be}{\begin{equation}}
\newcommand{\ee}{\end{equation}}

\newcommand{\bm}[1]{\mbox{\boldmath $#1$}}

\newcommand{\kslash}{k \!\!\! / }

\newcommand{\lslash}{l \!\! / }
\newcommand{\Pslash}{P \!\!\!\! / }

\newcommand{\islash}{i \!\!\! / }
\newcommand{\jslash}{j \!\!\! / }
\newcommand{\aslash}{a \!\!\! / }
\newcommand{\bslash}{{b \hspace{-6pt} \slash} }

\newcommand{\onslash}{1 \!\!\! / }
\newcommand{\twslash}{2 \!\!\!/ }
\newcommand{\thslash}{3 \!\!\!/ }
\newcommand{\foslash}{4 \!\!\! / }
\newcommand{\fislash}{5 \!\!\! / }

\newcommand{\mslash}{m \!\!\! / }

\def\bd{\begin{document}}
\def\ed{\end{document}}
\def\nn{\nonumber}
\def\bea{\begin{eqnarray}}
\def\eea{\end{eqnarray}}

\def\red{\color{red}}
\def\black{\color{black}}
\def\blue{\color{blue}}
\def\orange{\color{orange}}

\def\ab{(ijab)}
\def\ba{(ijba)}
\def\ijab{{\tr}_{+}(\islash\, \jslash\, \aslash \, \bslash)}
\def\ijba{{\tr}_{+}(\islash\, \jslash\, \bslash \, \aslash)}
\def\ijaP{{\tr}_{+}(\islash\, \jslash\, \aslash \, \Pslash)}
\def\ijPLa{{\tr}_{+}(\islash\, \jslash\, \Pslash_L \, \aslash)}
\def\ijaPL{{\tr}_{+}(\islash\, \jslash\, \aslash \, \Pslash_L)}
\def\ijPLza{{\tr}_{+}(\islash\, \jslash\, \Pslash_{L;z} \, \aslash)}
\def\ijaPLz{{\tr}_{+}(\islash\, \jslash\, \aslash \, \Pslash_{L;z})}
\def\ijPa{{\tr}_{+}(\islash\, \jslash\, \Pslash \, \aslash)}
\def\iaPb{{\tr}_{+}(\islash\, \aslash\, \Pslash \, \bslash)}
\def\ibPa{{\tr}_{+}(\islash\, \bslash\, \Pslash \, \aslash)}
\def\ijPmu{{\tr}_{+}(\islash\, \jslash\, \Pslash \, \mu)}
\def\ibmuP{{\tr}_{+}(\islash\, \bslash\, \mu \, \Pslash)}
\def\ibmua{{\tr}_{+}(\islash\, \bslash\, \mu \, \aslash)}
\def\iamub{{\tr}_{+}(\islash\, \aslash\, \mu \, \bslash)}
\def\jaPb{{\tr}_{+}(\jslash\, \aslash\, \Pslash \, \bslash)}
\def\ijmuP{{\tr}_{+}(\islash\, \jslash\, \mu \, \Pslash)}
\def\ijmum{{\tr}_{+}(\islash\, \jslash\, \mu \, \mslash)}
\def\ijmmu{{\tr}_{+}(\islash\, \jslash\, \mslash \, \mu)}
\def\ijmP{{\tr}_{+}(\islash\, \jslash\, \mslash \, \Pslash)}
\def\iabP{{\tr}_{+}(\islash\, \aslash\, \bslash \, \Pslash)}
\def\ijbP{{\tr}_{+}(\islash\, \jslash\, \bslash \, \Pslash)}
\def\jbPa{{\tr}_{+}(\jslash\, \bslash\, \Pslash \, \aslash)}
\def\ijPb{{\tr}_{+}(\islash\, \jslash\, \Pslash \, \bslash)}
\def\jbmua{{\tr}_{+}(\jslash\, \bslash\, \mu \, \aslash)}

\def\loablt{ {\tr}_{+}(\lslash_1\, \aslash \, \bslash\, \lslash_2)}

 \def\ijlolt{{\tr}_{+}(\islash\, \jslash\, \lslash_1 \, \lslash_2)}
\def\ijltlo{{\tr}_{+}(\islash\, \jslash\, \lslash_2 \, \lslash_1)}
\def\ibloa{{\tr}_{+}(\islash\, \bslash\, \lslash_1 \, \aslash)}
\def\jaltb{{\tr}_{+}(\jslash\, \aslash\, \lslash_2 \, \bslash)}
\def\ialtb{{\tr}_{+}(\islash\, \aslash\, \lslash_2 \, \bslash)}
\def\bltloa{{\tr}_{+}(\bslash\, \lslash_2\, \lslash_1 \, \aslash)}
\def\jbloa{{\tr}_{+}(\jslash\, \bslash\, \lslash_1 \, \aslash)}
\def\ibPb{{\tr}_{+}(\islash\, \bslash\, \Pslash \, \bslash)}
\def\ijltb{{\tr}_{+}(\islash\, \jslash\, \lslash_2 \, \bslash)}

\def\ijloa{{\tr}_{+}(\islash\, \jslash\,  \lslash_1 \, \aslash)}
\def\ijblt{{\tr}_{+}(\islash\, \jslash\,  \bslash \, \lslash_2)}

\def\jakb{{\tr}_{+}(\jslash\, \aslash\, \kslash \, \bslash)}
\def\iakb{{\tr}_{+}(\islash\, \aslash\, \kslash \, \bslash)}

\def\tofo{{\tr}_{+}(\onslash\, \thslash\, \twslash \, \foslash)}
\def\foto{{\tr}_{+}(\onslash\, \thslash\, \foslash \, \twslash)}
\def\tofi{{\tr}_{+}(\onslash\, \thslash\, \twslash \, \fislash)}
\def\fito{{\tr}_{+}(\onslash\, \thslash\, \fislash \, \twslash)}

\def\lrangle#1#2{\langle #1\,#2\rangle}

\def\Li{{\rm{Li}}}
\def\eps{\epsilon}
\def\epsuv{{\epsilon_{\rm \mbox{\tiny UV}}}}
\let\bm=\bibitem
\let\la=\label

\def\npb#1#2#3{Nucl. Phys. {\bf{B#1}} #3 (#2)}
\def\plb#1#2#3{Phys. Lett. {\bf{#1B}} #3 (#2)}
\def\prl#1#2#3{Phys. Rev. Lett. {\bf{#1}} #3 (#2)}
\def\prd#1#2#3{Phys. Rev. {D \bf{#1}} #3 (#2)}
\def\cmp#1#2#3{Comm. Math. Phys. {\bf{#1}} #3 (#2)}
\def\cqg#1#2#3{Class. Quantum Grav. {\bf{#1}} #3 (#2)}
\def\nppsa#1#2#3{Nucl. Phys. B (Proc. Suppl.) {\bf{#1A}}#3 (#2)}
\def\ap#1#2#3{Ann. of Phys. {\bf{#1}} #3 (#2)}
\def\ijmp#1#2#3{Int. J. Mod. Phys. {\bf{A#1}} #3 (#2)}
\def\rmp#1#2#3{Rev. Mod. Phys. {\bf{#1}} #3 (#2)}
\def\mpla#1#2#3{Mod. Phys. Lett. {\bf A#1} #3 (#2)}
\def\jhep#1#2#3{J. High Energy Phys. {\bf #1} #3 (#2)}
\def\atmp#1#2#3{Adv. Theor. Math. Phys. {\bf #1} #3 (#2)}
\newcommand{\EQ}[1]{\begin{equation} #1 \end{equation}}
\newcommand{\AL}[1]{\begin{subequations}\begin{align} #1 \end{align}\end{subequations}}
\newcommand{\SP}[1]{\begin{equation}\begin{split} #1 \end{split}\end{equation}}
\newcommand{\ALAT}[2]{\begin{subequations}\begin{alignat}{#1} #2 \end{alignat}
                        \end{subequations}}
\def\beqa{\begin{eqnarray}}
\def\eeqa{\end{eqnarray}}
\def\beq{\begin{equation}}
\def\eeq{\end{equation}}
\def\sst{\scriptscriptstyle}
\def\thetabar{\bar\theta}
\def\Tr{{\rm Tr}}
\def\one{\mbox{1 \kern-.59em {\rm l}}}
 \def\Nh{\hat{N}}

\newcommand{\half}{{\textstyle \frac{1}{2}}}

\def\a{\alpha}      \def\da{{\dot\alpha}}
\def\b{\beta}       \def\db{{\dot\beta}}
\def\c{\gamma}  \def\G{\Gamma}  \def\cdt{\dot\gamma}
\def\d{\delta}  \def\D{\Delta}  \def\ddt{\dot\delta}
\def\e{\epsilon}        \def\vare{\varepsilon}
\def\f{\phi}    \def\F{\Phi}    \def\vvf{\f}
\def\h{\eta}
\def\k{\kappa}
\def\l{\lambda} \def\L{\Lambda}
\def\m{\mu} \def\n{\nu}
\def\o{\omega}
\def\p{\pi} \def\P{\Pi}
\def\r{\rho}
\def\s{\sigma}  \def\S{\Sigma}
\def\t{\tau}
\def\th{\theta} \def\Th{\Theta} \def\vth{\vartheta}
\def\X{\Xeta}
\def\z{\zeta}
\def\de{\partial}
\def\cA{{\cal A}} \def\cB{{\cal B}} \def\cC{{\cal C}}
\def\cD{{\cal D}} \def\cE{{\cal E}} \def\cF{{\cal F}}
\def\cG{{\cal G}} \def\cH{{\cal H}} \def\cI{{\cal I}}
\def\cJ{{\cal J}} \def\cK{{\cal K}} \def\cL{{\cal L}}
\def\cM{{\cal M}} \def\cN{{\cal N}} \def\cO{{\cal O}}
\def\cP{{\cal P}} \def\cQ{{\cal Q}} \def\cR{{\cal R}}
\def\cS{{\cal S}} \def\cT{{\cal T}} \def\cU{{\cal U}}
\def\cV{{\cal V}} \def\cW{{\cal W}} \def\cX{{\cal X}}
\def\cY{{\cal Y}} \def\cZ{{\cal Z}}
\def\ua{\underline{\alpha}}
\def\ub{\underline{\phantom{\alpha}}\!\!\!\beta}
\def\uc{\underline{\phantom{\alpha}}\!\!\!\gamma}
\def\um{\underline{\mu}}
\def\ud{\underline\delta}
\def\ue{\underline\epsilon}
\def\una{\underline a}\def\unA{\underline A}
\def\unb{\underline b}\def\unB{\underline B}
\def\unc{\underline c}\def\unC{\underline C}
\def\und{\underline d}\def\unD{\underline D}
\def\une{\underline e}\def\unE{\underline E}
\def\unf{\underline{\phantom{e}}\!\!\!\! f}\def\unF{\underline F}
\def\unm{\underline m}\def\unM{\underline M}
\def\unn{\underline n}\def\unN{\underline N}
\def\unp{\underline{\phantom{a}}\!\!\! p}\def\unP{\underline P}
\def\unq{\underline{\phantom{a}}\!\!\! q}
\def\unQ{\underline{\phantom{A}}\!\!\!\! Q}
\def\unH{\underline{H}}
\def\As {{A \hspace{-6.4pt} \slash}\;}
\def\bs {{b \hspace{-6.4pt} \slash}\;}
\def\Ds {{D \hspace{-6.4pt} \slash}\;}
\def\ds {{\del \hspace{-6.4pt} \slash}\;}
\def\ss {{\s \hspace{-6.4pt} \slash}\;}
\def\ks {{ k \hspace{-6.4pt} \slash}\;}
\def\ps {{p \hspace{-6.4pt} \slash}\;}
\def\pas {{{p_1} \hspace{-6.4pt} \slash}\;}
\def\pbs {{{p_2} \hspace{-6.4pt} \slash}\;}
\def\Ps {{P \hspace{-6.4pt} \slash}\;}
\def\Qs {{Q \hspace{-6.4pt} \slash}\;}
\def\Fh{\hat{F}}
\def\Vh{\hat{V}}
\def\Xh{\hat{X}}
\def\ah{\hat{a}}
\def\xh{\hat{x}}
\def\yh{\hat{y}}
\def\ph{\hat{p}}
\def\xih{\hat{\xi}}
\def\psit{\tilde{\psi}}
\def\Psit{\tilde{\Psi}}
\def\tht{\tilde{\th}}
\def\lt{\tilde{\lambda}}
\def\hl{\hat{\lambda}}
\def\hlt{\hat{\tilde{\lambda}}}
\def\llt{\tilde{l}}
\def\At{\tilde{A}}
\def\Qt{\tilde{Q}}
\def\Rt{\tilde{R}}
\def\Nt{\tilde{N}}

\def\at{\tilde{a}}
\def\st{\tilde{s}}
\def\ft{\tilde{f}}
\def\pt{\tilde{p}}
\def\qt{\tilde{q}}
\def\vt{\tilde{v}}
\def\nt{\tilde{n}}
\def\delb{\bar{\partial}}
\def\bz{\bar{z}}
\def\bD{\bar{D}}
\def\bB{\bar{B}}
\def\bk{{\bf k}}
\def\bl{{\bf l}}
\def\bp{{\bf p}}
\def\bq{{\bf q}}
\def\br{{\bf r}}
\def\bx{{\bf x}}
\def\by{{\bf y}}
\def\bR{{\bf R}}
\def\bV{{\bf V}}
\def\d{\delta}\def\D{\Delta}\def\ddt{\dot\delta}
\def\pa{\partial} \def\del{\partial}
\def\xx{\times}
\def\uno{\mbox{1 \kern-.59em {\rm l}}}
\def\trp{^{\top}}
\def\inv{^{-1}}
\def\dag{{^{\dagger}}}
\def\pr{^{\prime}}
\def\lan{\langle}
\def\ran{\rangle}
\def\rar{\rightarrow}
\def\lar{\leftarrow}
\def\lrar{\leftrightarrow}
\newcommand{\0}{\,\!}      %this is just NOTHING!
\def\one{1\!\!1\,\,}
\def\im{\imath}
\def\jm{\jmath}
\newcommand{\tr}{\mbox{tr}}
\newcommand{\slsh}[1]{/ \!\!\!\! #1}
\def\vac{|0\rangle}
\def\lvac{\langle 0|}
\def\hlf{\frac{1}{2}}
\def\ove#1{\frac{1}{#1}}
\def\Box{\square}
\def\ZZ{\mathbb{Z}}
\def\CC#1{({\bf #1})}
\def\bcomment#1{}
\def\bfhat#1{{\bf \hat{#1}}}
\def\VEV#1{\left\langle #1\right\rangle}
\newcommand{\ex}[1]{{\rm e}^{#1}} \def\ii{{\rm i}}
\def\rr{{\rm r}} \def\rs{{\rm s}}\def\rv{{\rm v}}
\def\ri{{\rm i}}\def\rj{{\rm j}}
\newcommand{\lrbrk}[1]{\left(#1\right)}
\newcommand{\sfrac}[2]{{\textstyle\frac{#1}{#2}}}

\font\mybb=msbm10 at 12pt
\def\bb#1{\hbox{\mybb#1}}

\font\myBB=msbm10 at 18pt
\def\BB#1{\hbox{\myBB#1}}

\begin{document}

\begin{flushright}
IPPP/11/49, DCPT/11/98
\end{flushright}

\vspace{6pt}

\begin{center}

 \hspace{-0.8cm}{\Large \bf Wilson Loops @ 3-Loops in Special Kinematics}

\vspace{15pt}
\end{center}

\centerline{\mbox {\large Paul~Heslop$^{a}$ and
Valentin~V.~Khoze$^{b}$}%
{
\renewcommand{\thefootnote}{}  \footnotetext{
{\tt paul.heslop@durham.ac.uk, valya.khoze@durham.ac.uk} } } }

\begin{center}
{\small \em
\begin{itemize}
\item[\ \ \ \ \ \ $^a$]
Department of Mathematical Sciences\\
Durham University,
Durham, DH1 3LE, United Kingdom\\
\item[\ \ \ \ \ \ $^b$]
Institute for Particle Physics Phenomenology,  \\
Department of Physics,
Durham University, \\
Durham, DH1 3LE, United Kingdom

\end{itemize}
}

%\vspace{-8pt}

\vspace{23pt} {\bf Abstract}
\end{center}

\noindent
We obtain a compact expression for the octagon MHV amplitude / Wilson loop at 3
loops in planar $\cN$=4 SYM and in special 2d kinematics in terms of 7
unfixed coefficients. We do this by making use of the cyclic and parity symmetry
of the amplitude/Wilson loop and its behaviour in the soft/collinear
limits as well as in the leading term in the expansion away from this
limit. We also make 
a natural and quite general assumption about the functional form of
the result, namely that it should consist of weight 6 polylogarithms
whose symbol consists of basic cross-ratios only (and not
functions thereof). We also describe the uplift of this result to 10 points.

\setcounter{page}{0} \thispagestyle{empty}
\newpage

%%%%%%%%%%%%%%%%%%%%%%%%%%%%%%%%%%%%%%%%%%%%%%%%%%%%%%%%%%%%%%%%%
\def\cS{{\cal S}}
\def \tens{\otimes}

\section{Introduction}
\setcounter{footnote}{0}

In~\cite{2loop} an infinite  sequence of MHV amplitudes / Wilson loops in a special kinematical
regime was found. The sequence  started with  the first non-trivial amplitude in
the sequence, the 8-point case obtained  by direct computation
in~\cite{dds8}, but the higher point results
were obtained by using a simple
assumption concerning the structure of the result, together with
collinear limits and cyclic and parity symmetry.
 In this paper we wish
to push these ideas to 3-loops.

Our understanding of perturbative scattering amplitudes in planar $\cN=4$ SYM is currently
increasing at a rather  rapid rate. Indeed in just the last year or
so the fruitful duality between MHV amplitudes and
Wilson
loops~\cite{am,dks,bht} has been formally
extended to arbitrary amplitudes~\cite{marksAndSparks,CaronHuot:2010ek}
once issues of regularisation are properly understood~\cite{belitsky}.
A new duality between correlation
functions and both Wilson loops and amplitudes has been
found~\cite{AEMKS,EKS2,EKS3,paper1,mscorrelationfunctions,paper2} and
this has already
proved useful in both directions,  obtaining previously unknown
correlation functions
using known amplitudes as well as providing new insights into amplitudes themselves~\cite{Eden:2011we}.
And a loop-level integrand
version~\cite{Boels:2010nw,ArkaniHamed:2010kv} of
the BCFW recursion relation~\cite{Britto:2005fq} has enabled one to find
arbitrary loop level amplitude integrands
from purely algebraic
methods~\cite{ArkaniHamed:2010kv,ArkaniHamed:2010gh}.

The above  impressive results have been largely formulated
at the level of the integrand. Of course ultimately we are interested
in the amplitudes themselves, the result of having performed the
integration of these integrands. Much progress has also been made
here but
as yet
at a somewhat more modest level, and most of the
developments~\cite{dhks4,dhks5,seven,dhks6,Anastasiou:2009kn,6an1,6an2,dds8,2loop,CaronHuot:2011ky,Dixon:2011pw}
are still driven by the original MHV amplitude/
Wilson loop duality, and result from the fact that the Wilson loop
integrals are simpler
than the amplitude ones.
A major new mathematical tool, arising from this is the notion of
the ``symbol''~\cite{gonch,Goncharov:2010jf}. This allows one to map
highly complicated
polylogarithmic functions to tensors involving rational
functions. In this way obscure polylogarithmic identities become manifest
algebraic identities satisfied by this tensor. This allowed the
authors of~\cite{Goncharov:2010jf} to reduce the huge formula arising
from the impressive direct computation of the
hexagon Wilson loop at two-loops~\cite{6an1,6an2} to a single
line~\cite{Goncharov:2010jf}. Indeed the most recent results
concerning amplitudes at the integr{\em{\!al}} level  have actually
been given as
symbols rather than the functions
themselves~\cite{CaronHuot:2010ek,Dixon:2011pw}. For example, very recently
the symbol of the  3-loop hexagon Wilson loop was derived up to two
unfixed coefficients in~\cite{Dixon:2011pw}.

Another new tool for analytic amplitude computations is the
OPE/near collinear limit~\cite{Alday:2010ku,gmsv,Gaiotto:2011dt,Sever:2011da}
allowing an expansion around the collinear limit to be understood
in terms of an OPE expansion. At the moment there is an
obstruction to going beyond the next to leading term in this
expansion, but even at this level we obtain important information about the
amplitude which we will make use of here.

In order to investigate further perturbative amplitudes
without doing a direct
computation, we will
restrict ourselves to
the so-called $AdS_3$ special kinematics, first introduced
in~\cite{am8} in the strong coupling context. This corresponds to
assuming that all the external momenta live in $1+1$ dimensions rather
than the full $3+1$ dimensions. These provide a nice arena for
studying non-trivial high loop order amplitudes/Wilson loops whilst
avoiding some of the kinematical complications of the full amplitudes.

In~\cite{2loop} we were able to take the 2-loop result for the 8-point Wilson loop in
special kinematics,
computed directly in~\cite{dds8} and extend it to all (even)
$n$-points, using symmetry and collinear limits as well as a simple
assumption about its structure. The assumption was that at 2-loops the conformal part of the answer
should depend only on logarithms
of $x$ space cross-ratios
\begin{equation}\label{uijdef}
  u_{ij}={x_{ij+1}^2 x^2_{i+1 j} \over x_{ij}^2 x_{i+1 j+1}^2} \ .
\end{equation}
We then verified that our analytic expressions for all $n$ agreed with numerical computations
carried out following the numerical algorithm developed in~\cite{Anastasiou:2009kn} and further used in \cite{bhkt}.

However the $\log$s-only structure of the answer cannot be expected to hold  beyond
2-loops since the OPE implies the presence of polylogarithms at 3 loop level~\cite{Alday:2010ku}.
The crucial insight which enables us
to go further in this sector then, is our expectation that, despite the known
complicated variables which occur in MHV amplitudes at two-loops and
beyond for general kinematics, we expect that these all simplify in special
kinematics. Indeed all expected variables in general
kinematics
(for example those given in~\cite{DelDuca:2011wh}) reduce to simple
cross-ratios.
So we will assume in this paper that the symbol
takes values only over the standard $x$-space
cross-ratios~(\ref{uijdef}). In other words we relax the assumption we
made at two loops that the amplitude depends only on logarithms, but
we maintain the assumption that the arguments of the symbol should be simple
cross-ratios only.

So then using this assumption together with cyclic and parity symmetry
of the Wilson loop/MHV amplitude, and the important restriction that
the symbol should arise from a function (the so called integrability constraint)
we can firstly derive the 8 point 2-loop result of~\cite{dds8} without
computation (with one unfixed coefficient), and prove that the
uplift to $n$-points found in~\cite{2loop} is in fact the unique
solution of our constraints. At 3-loops we can
restrict the 3-loop 8-point amplitude down to just 13 unfixed
coefficients. The further constraints arising from the OPE/collinear
limit then reduces this to 7 unfixed coefficients.
At higher $n$ we are able to uplift the result to 10 points,
albeit with the introduction of 12 new unfixed coefficients. The
uplift to 12 points can also be performed, but again there
will be further new unfixed coefficients introduced. However, the uplift from
12 points to 14 points and beyond  is then unique at 3 loops within our ansatz.

More generally, at $l$ loops, once the $4l$-point function is known
the uplift via inverse soft/triple collinear limits is unique.

Although we initially perform all this analysis at the level of the symbol, we
are able to invert the symbol
and obtain the functions themselves. Indeed although the
corresponding symbols
become quickly very large indeed with increasing $n$, the functions
themselves can be written fairly compactly.

In section~\ref{sec:background-material} we review the set up and
some background material we will need. In section~\ref{sec:fud-assum}
we discuss further our assumption that only $u_{ij}$'s should appear
in the symbol. Section~\ref{sec:one-two-loop} reviews the remainder
function at one- and two-loops from this
perspective. In section~\ref{sec:more-syst-appr} we determine  the octagon
3-loop amplitude as far as we can and in
section~\ref{sec:lift-high-points} we discuss the uplift to higher
points at 3-loops.

\section{Background material}

MHV amplitudes and null polygonal Wilson loops in planar $\cN=4$ SYM are traditionally characterised by the remainder function
${\cR}_n$ which is defined as the
difference between the logarithm of the Wilson loop $W_n$ and the known BDS expression of~\cite{abdk,bds},
\begin{equation}\label{eq:1111}
  {\cR}_n \,=\,\log (W_n) -(BDS)^{WL}_n\ .
\end{equation}
${\cR}_n$ is a conformally-invariant function and thus depends  only
on
conformally-invariant cross-ratios \cite{dks,dhks5}.
For a polygonal contour with $n$ light-like edges, in general,
there are $n(n-5)/2$ independent conformal cross-ratios (if we do not,
as in \cite{Anastasiou:2009kn}, impose the Gram determinant
constraints). A basis for the
cross ratios is provided by $u_{ij}$, defined
in~(\ref{uijdef}).

\label{sec:background-material}
\subsection{Special kinematics}

In this paper we will be restricting our attention exclusively to the
case of special kinematics,
first introduced in \cite{am8},
 where the external momenta lie entirely in
$1+1$ dimensions.
For the Wilson loop contour to be embeddable into two space-time
dimensions the number of edges $n$ must be even and the number of independent cross-ratios reduces and they
have to satisfy the following conditions,
\begin{eqnarray}
u_{i\,,  i+ {\rm odd}} &=& 1 \nonumber \\
u_{2i+1\,,   2j+1} &=& u^+_{ij} \qquad    2 \leq (i-j)\, \, {\rm mod} \, n/2  \le n/2-2    \label{eq:AdS3} \\
u_{2i\,,  2j} &=& u^-_{ij}\nonumber
\end{eqnarray}
Here
the vertices of the contour have the
following simple light-cone representation:
\begin{align}
  x_{2i}=(x_i^+,x_i^-)\ , \qquad   x_{2i+1}=(x_i^+,x_{i+1}^-)\ , \qquad i=1,\ldots, n \,,
\end{align}
and
the cross-ratios $u_{ij}^{\pm}$ appearing on the right hand side
of \eqref{eq:AdS3} are functions of only either $x^+$ or $x^-$
light-cone coordinates:
\begin{equation}\label{uijdef2}
  u_{ij}^{+}\, :=\,{x^+_{ij+1}\, x^+_{i+1 j} \over x^+_{ij}\, x^+_{i+1 j+1}} \ , \qquad
   u_{ij}^{-}\, :=\,{x^-_{ij+1}\, x^-_{i+1 j} \over x^-_{ij}\, x^-_{i+1 j+1}} \ .
\end{equation}
As such, these cross-ratios are essentially made from
one-dimensional distances.  This results
in the following simple identity
\begin{align}
\label{ueq}
  (1-u^\pm_{i\, j+1})   (1-u^\pm_{i+1\, j})& \,=\,   (1- 1/{u^\pm_{i\,
      j}})  (1-1/{u^\pm_{i+1\, j+1}})\\
       u_{i, i+1}=u_{i+1, i}&=0 \qquad u_{i,i }=\infty  \ ,
\end{align}
which  is precisely  the $AdS_3$ Y-system equation
of~\cite{agm}, where the $Y$'s of~\cite{agm} (evaluated at
$\zeta=0$) are associated with the cross-ratios as
\begin{align}
  u_{k,-k-1}^+ = {Y_{2k} \over 1+Y_{2k}} \qquad \qquad u_{k,-k-2}^- =
  {Y_{2k+1} \over 1+Y_{2k+1}}\ .
\end{align}
We will thus refer to~(\ref{ueq}) as the Y-system from now on.

For the two lowest-$n$ cases, the octagon and the decagon, all the cross-ratios different from 1 in \eqref{eq:AdS3}
are of the form $u_{i,i+4}$, with $i=1,\ldots,4$ for the octagon, and $i=1,\ldots,10$ for the decagon.
To simplify notation in these two cases, we define $u_i:= u_{i,i+4}$. Similarly, for their
decomposition into $\pm$ components, we will often use $u^\pm_j:= u^\pm_{j,j+2}$.

Clearly, the cross-ratios $u_i$ are not all independent, as we have seen above, they are further constrained by the Y-system
equations, leaving $n-6$ (i.e. 2 for the octagon and 4 for the decagon) independent solutions. Nevertheless, as in our earlier work,
\cite{2loop} we will use the full set of $u_i$ as the set of variables appearing in all expressions.

More details of the special kinematics in this context can be found
in~\cite{2loop}.

\subsection{Collinear limits}
\label{sec:collinear-limits}
The collinear limits which allow us to remain in the special
kinematics have to reduce the number of edges (number of external
momenta for amplitudes) by an even number. The minimal such limit is
the triple-collinear limit in which three consecutive edges become collinear.\footnote{A more
appropriate way to visualise this limit in the way which is consistent with the zig-zag construction
of the polygon, is in terms of the collinear-soft-collinear limit. In this case the middle edge becomes
soft and the two edges, one on the left and one on the right of it, are collinear to each other; thus the three edges
are reduced to one.}
For concreteness, consider the limit of $\cR_n$ in which edges
$n - 2$, $n - 1$ and $n$
become collinear (and in which in fact edge $n -1$ becomes soft). In this case one has
\begin{equation}
u_{i,n-1} \rightarrow 1\, , \qquad u_{1,n-3} \rightarrow 0\, ,
\label{coll-lim}
\end{equation}
while the remaining cross-ratios $u_{i,j}$ remain unchanged.
The remainder function $\cR_n$ reduces in this limit to $\cR_{n-2}$ plus a correction $\cR_6$ arising from the
triple-collinear splitting function (in our special kinematics $\cR_6={\rm const}$). Specifically, one has
\cite{Anastasiou:2009kn,2loop}
\begin{equation}
\cR_n(u_{i,j}) \rightarrow \cR_{n-2}(\hat{u}_{i,j})+ \cR_6\, .
\label{R-split}
\end{equation}
Here the $(n -2)$-point cross-ratios $\hat{u}_{i,j}$ are defined in terms of the
$n$-point cross-ratios in the collinear limit as
\begin{equation}
\hat{u}_{i,n-2} = u_{i,n-2} \, u_{i,n} \, , \qquad \hat{u}_{i,j} = u_{i,j} \quad i,j \neq n -2
\, .
\label{uhat-split}
\end{equation}
In particular, for the octagon we have $\cR_8(u_i) \rightarrow 2 \, \cR_6 = {\rm const},$ and for the
decagon,
\begin{equation}
\cR_{10}(u_{i}) \rightarrow \cR_{8}(\hat{u}_{i})+ \cR_6 \quad {\rm where} \quad \hat{u}_{4}=\,u_4 \,u_{10}\, .
\label{R-split-2}
\end{equation}
In the above equation the $u$'s solve the 10-point Y-system equation~(\ref{ueq}) and the $\hat u$'s then automatically solve the 8-point Y-system equation.
From now on we will always refer to these (triple) collinear-soft limits as collinear limits.
For more detail on collinear limits in special kinematics we refer the reader to \cite{2loop}.

\subsection{Symbols}

The ``symbol'' is an important  new mathematical tool, introduced in the context
of particle physics in~\cite{Goncharov:2010jf},
and already  proving highly useful in
$\cN$=4 SYM amplitudes,
but which should also be
relevant more generally in particle physics (see for
example~\cite{Buehler:2011ev}).

The symbol associates to any (generalised) polylogarithm, a tensor
whose entries are rational functions of the arguments. The rank of the tensor is equal to
the weight of the polylogarithm. For example   $\log x$ has weight 1 and
gives rise to a 1-tensor
\begin{align}\label{eq:11}
  \cS\Big( \log x \Big) = x
\end{align}
whereas the classical polylogarithms  have symbol given as
\begin{align}\label{eq:16}
  \cS\Big(\Li_w(x)\Big) &= - (1-x)\tens \overbrace{x \tens \dots \tens
    x}^{w-1}\ .
\end{align}
The symbol has the following properties inherited from the logarithm
\begin{eqnarray}\label{eq:19}
  \dots \tens x\, y \tens \dots &=&\dots \tens x \tens \dots\, +\,  \dots
  \tens y \tens \dots \\
   \dots \tens 1/x \tens \dots &=& -\, \dots \tens x \tens \dots \nonumber
\end{eqnarray}
\black
from which follows the important property that the symbol vanishes when
any entry equals unity
\begin{align}
  \dots \tens 1 \tens \dots = 0\ .
\end{align}
It is also blind to multiplication by constants.
The final property of the symbol we need is the symbol of products of
functions. This is given by taking the shuffle product of the symbol
of each function
\begin{align}
  \cS (f g)= \cS (f) \shuffle \cS(g)\ .
\end{align}
For example
\begin{align}
  \cS( \Li_2(x) \log y) &= \Big(- (1-x) \tens x \Big) \shuffle y
  \nonumber \\
&= -
  (1-x) \tens x \tens y  -
  (1-x) \tens y \tens x  -
  y \tens (1-x) \tens x \, ,
\end{align}
or for three log functions we have,
\begin{align}
  &\cS(\log(x) \log (y) \log(z)) =\, x \shuffle y \shuffle z =\, \Big(x \tens y + y \tens x \Big) \shuffle z
  \nonumber \\
&= x \tens y \tens z + x \tens z \tens y +  z \tens x \tens y + y \tens x \tens z + y \tens z \tens x + z \tens y \tens x \, .
\end{align}

The symbol can be defined recursively. One can write
the total derivative of any weight $w$ generalised polylogarithm
(here by this we mean any function with a well-defined rank-$w$
symbol) as follows
\begin{align}\label{eq:17}
  d f = \sum_i g_i \, d \log(x_i)
\end{align}
where the $g_i$ are weight $w-1$ polylogarithms. Then the
corresponding symbol is given as
\begin{align}\label{eq:18}
  \cS \big( f \big) = \sum_i \cS\big(g_i\big) \tens x_i\ .
\end{align}
This definition (together with~(\ref{eq:11})) gives all the above properties.

The symbol is incredibly useful since it trivialises otherwise complicated
identities involving polylogarithms. The most spectacular example of
such a simplification
is the reduction of the formula found for the hexagon two-loop
Wilson loop in~\cite{6an1,6an2}  to the single line formula
in~\cite{Goncharov:2010jf}. However the inverse process of finding the
function from the symbol is far from straightforward to do in
practice.
Indeed the symbol  is often much more complicated and longer than the
actual functions which produce it due to the shuffle product for example.
The symbol is also
non-unique. It is equivalent to  the ``maximally transcendental'' piece of the
function, but all information about lower weight terms is lost in the
symbol.

The great advantage of the special kinematics we consider here is that
the functions that occur
will turn out to be relatively simple and after obtaining the symbol we will be able
to reconstruct the functional form in section~\ref{sec:more-syst-appr}.

\subsection{The integrability constraint}
\label{sec:integr}

The fact that $d^2 f=0$ together with its recursive
definition~(\ref{eq:17},\ref{eq:18}) give non-trivial and powerful constraints on symbols of
functions. Namely for a weight $w$ tensor we obtain the $w-1$ equations
\begin{align}
\label{eq:integr}
 & \cS(f) = \sum x_1 \tens \dots \tens x_w \\
 \Rightarrow \  &\sum d \log x_i \wedge d \log x_{i+1}
  \ \ x_1 \tens \dots \tens x_{i-1} \tens x_{i+2} \tens \dots \tens
  x_w = 0 \ .\nonumber
\end{align}
where there is no sum over $i$.
We will make extensive use of this constraint in deriving the 8-point
3-loop remainder function.

\section{Fundamental assumption: the symbol contains u's only}
\label{sec:fud-assum}

In the rest of this paper we will attempt to constrain, as far as
possible, the analytic form of the remainder functions using
symmetries and collinear limits. In order to do this we make one
fundamental assumption which makes this possible. Namely we assume
that the function has a symbol whose entries can always
be taken from the basis of cross-ratios $u_{i\,j}$.
In other words, the symbol is made of sums of the tensor products of $u_{i\,j}$'s, and no
functions of the cross-ratios should appear in the symbol.
This is certainly not the case in general kinematics
where, for example at 6-points  one
can have entries  $1-u$ as well as functions involving
square roots of combinations of $u$'s. However in the special
kinematics we consider,  we expect these will always reduce to $u$'s.

For example twistor brackets, in terms of which remainder function
symbols seem to be naturally given (see for example~\cite{ArkaniHamed:2010kv,Goncharov:2010jf,DelDuca:2011wh,Spradlin:2011wp})
always reduce in special kinematics to simple products of $x$'s. So
for example in a conformally invariant expression a four-bracket of
two even and two odd twistors
reduces as
\begin{align}
\vev{ 2i 2j (2k-1) (2l-1)} \rightarrow x_{ij}^+ x^-_{kl} \ ,
\end{align}
with any other possibility vanishing, whereas more complicated
twistor invariants which should
appear reduce
similarly, eg
\begin{align}
  \vev{X_{2i}  \bar Z_{2j} \cap \bar Z_{2k-1} } \sim x_{ij}^+
  \, x^-_{j\,j+1} \, x^+_{k-1\,k} \, x^-_{ik}\ .
\end{align}

Furthermore it is always possible to rewrite $1-u$ in terms of
product of  $u$'s
using the Y-system equations \eqref{ueq}.
Indeed one can check that
\begin{align}
\label{eq:Ysoln}
  1-u_{ij}^\pm = \prod_{k=i+1}^{j-1} \prod_{l=j+1}^{i-1} u_{kl}^\pm \ .
\end{align}
Clearly inside a symbol, using~(\ref{eq:19}),  this can then be written in terms of
a sum of terms involving $u_{ij}$'s.

For the case of octagon, \eqref{eq:Ysoln} collapses to
\begin{equation}
\label{eq:8Ysoln}
n=8\,: \qquad 1-u_1=u_3 \, , \qquad 1-u_2=u_4 \, ,
\end{equation}
and for the decagon we have
\begin{equation}
\label{eq:10Ysoln}
n=10\,: \qquad 1-u_i^\pm=u_{i-1}^\pm u_{i+1}^\pm \, \quad i=1,\ldots,5 \, .
\end{equation}

In summary the natural assumption that the entries in the symbol are always
$u_{ij}$'s in special kinematics, is consistent with all expectations
for general kinematics.

\section{The one- and two-loop remainder functions revisited}
\label{sec:one-two-loop}

In this section we revisit the two-loop $n$-point remainder
functions in $1+1$ dimensional kinematics. The 8-point remainder
function was first obtained by a direct computation of the Wilson
loop in~\cite{dds8}. It can
be written as
\begin{align}\label{r8}
  \cR_{8}^{(2)}= -\frac{1}{2} \log(u_1) \log(u_2) \log(u_3) \log(u_4)
\, -\, \frac{\pi^4}{18}\ .
\end{align}
where $u_i:=u_{i i+4}$. We then  uplifted this in~\cite{2loop} to give
the two-loop remainder
function for any  $n$ (in 1+1 dimensions) in the remarkably concise form
\begin{align}\label{rn}
\cR_n^{(2)}
&= -{1 \over 2} \Big( \sum_{\cS} \log ( u_{i_1
    i_5}) \log ( u_{i_2 i_6}) \log ( u_{i_3 i_7}) \log ( u_{i_4 i_8})
  \Big) - {\pi^4\over 72} (n-4)\ ,
\end{align}
where the sum runs over the set
\begin{align}
\cS &= \Big\{ i_1, \dots i_8: 1\leq i_1<i_2< \dots < i_8 \leq n,\qquad  i_k
-i_{k-1} = \mathrm{odd} \Big\}\ .
\end{align}

This uplift from 8 points to $n$ points
was done
by considering collinear limits alone. We found
functions satisfying these and we then
checked the result using the numerical code constructed
in~\cite{Anastasiou:2009kn}.

This result was derived in \cite{2loop} following the assumption (based on the explicit form of the 8-point 2-loop result
as well as the $n$-point 1-loop results)  that only $\log$s of cross-ratios can appear.
This is correct at 2-loops, but at 3-loops the
OPE analysis suggests that one needs to
consider more general functions than simple logarithms~\cite{gmsv}. To find the strategy which works at all loops,
we are thus lead to re-derive the two-loop results \eqref{r8} and \eqref{rn} from a weaker assumption.
In this
paper therefore we will instead make the much less restrictive
assumption (motivated in section~\ref{sec:fud-assum}) that the function has a
symbol as a sum of tensor products of basis cross-ratios,
$u_{ij}$'s. As we shall see, this weaker assumption, together with
collinear limits, and cyclic and parity symmetry implies the
appearance of $\log(u)$'s only at two loops.

It turns out that under this simple and natural assumption,  we
can both rule out the existence of a 1 loop remainder function and
derive, without any direct computations, the 8-point 2-loop remainder
function (up to 1 unfixed constant). We will also show that under this
assumption the uplift to
the $n$-point 2-loop remainder (found in~\cite{2loop}) is unique (but not the 3-loop uplift which will be constructed
in a later section).

\subsection{Non-existence of a 1 loop 8-point remainder}
\label{sec:1-loop-remainder}

The $n$-point remainder function at any loop order must reduce under the
collinear limit to the $n-2$-point remainder function plus the
$6$-point remainder (which is a constant
in the $1+1$ dimensional kinematics). So we can consider
\begin{align}
 \tilde \cR_n= \cR_n - \half(n-4)\cR_6
\end{align}
which simply reduces as  $\tilde \cR_n \rightarrow \tilde \cR_{n-2}$ in the collinear
limit. In particular $\tilde \cR_{6}=0$ and so $\tilde
\cR_{8} \rightarrow 0$ in the collinear limit.

Now at 1 loop one can quickly see that there is no weight-2 symbol (ie no 2-tensor)
we can write down which will vanish in all collinear
limits. The collinear limit is $u_4 \rightarrow 1$, $u_2 \rightarrow
0$ but with $u_1=1-u_3$ left arbitrary. So in order for a tensor
involving $u$'s only to vanish in this collinear limit, all terms must
therefore contain a $u_4$. But cyclic symmetry ensures that this can
never be the case. We therefore immediately rule out a 1-loop 8-point
collinear vanishing remainder.

Similar considerations rule out the 1-loop $n$-point remainder function.

\subsection{Uniqueness of the 2-loop 8-point remainder}
\label{sec:uniqueness-2-loop}

Let us now consider therefore the most general possible collinear vanishing
2-loop remainder  function. This will give a nice illustration of the
technique we will implement later in more general cases.

We wish to write down the most general 2-loop (ie weight 4) symbol
which has dihedral symmetry (cyclic  + parity) and vanishes in any
soft/triple collinear limit. In order for the symbol to vanish in any
collinear limit, each term in the symbol must contain all four
cross-ratios $u_1, u_2, u_3, u_4$. Indeed if a term contains just 3
out of the four $u$'s, then this will never vanish under the
particular collinear
limit which has the remaining $u\rightarrow 1$. For example if
one chooses
a term to be $u_1\tens u_1 \tens u_2 \tens u_3$ then under the
collinear limit $u_4 \rightarrow 1$, $u_2 \rightarrow 0$, this will
not vanish (indeed it will
diverge). Furthermore this can never be compensated by a similar
non-vanishing term in the symbol. We therefore  consider  all 4!
terms in the symbol which contain all 4 cross-ratios, $u_1 \tens u_2
\tens u_3 \tens u_4$ together with permutations. Now we impose
dihedral symmetry generated by
\begin{align}\label{eq:20}
  u_1\rightarrow u_2 \rightarrow u_3 \rightarrow u_4 \rightarrow u_1
  \quad \mbox{and} \quad u_1\leftrightarrow u_4,\    u_2
  \leftrightarrow u_3\ .
\end{align}
In this way we obtain just three independent symbols
\begin{align}
\cR_8^{(2)}&= a \cR_{8;a}^{(2)} + b \cR_{8;b}^{(2)}+c \cR_{8;c}^{(2)}+
2 \cR_6^{(2)}\label{eq:10}\\[10pt]
 \cS\Big(\cR_{8;a}^{(2)}\Big)&= u_1 \tens u_2 \tens u_3 \tens u_4  \ + \ \mbox{ 7 terms related by
    dihedral symmetry}\nonumber\\
 \cS\Big(\cR_{8;b}^{(2)}\Big)&= u_1 \tens u_2 \tens u_4 \tens u_3  \ + \ \mbox{ 7 terms related by
    dihedral symmetry}\nonumber\\
 \cS\Big(\cR_{8;c}^{(2)}\Big)&= u_1 \tens u_3 \tens u_2 \tens u_4  \ + \ \mbox{ 7 terms related by
    dihedral symmetry}\ .
\end{align}
All three terms separately vanish in the collinear limit, and are symmetric
under the full dihedral symmetry. However they are {\em{not} }
necessarily symbols of functions. The integrability constraint, $d^2
\cR_8^{(2)}=0$ imposes
constraints on the allowed symbols
as described in section~\ref{sec:integr}.
We get three equations from the
derivatives hitting the first two entries, the second and third entries or
the third and fourth entries respectively in the symbol:
\begin{align}\nonumber
  a\,  {d u_1 \wedge d u_2 \over u_1 u_2}  \,u_3 \tens u_4  \ + \ b \,{d u_1 \wedge
  d u_2  \over u_1 u_2}\, u_4 \tens u_3  \ + \ c \,{du_1 \wedge du_3  \over u_1 u_3}\,  u_2
  \tens u_4  \ + \ \mbox{  dihedral } \ =\ &0\\\nonumber
  a \,{du_2 \wedge du_3  \over u_2 u_3}\,  u_1 \tens u_4  \ + \ b \,{du_2 \wedge
  du_4  \over u_2 u_4}\,  u_1 \tens u_3  \ + \ c \,{du_3 \wedge du_2  \over u_2 u_3}\,  u_1
  \tens u_4  \ + \ \mbox{  dihedral } \ =\ &0\\
  a \,{du_3 \wedge du_4  \over u_3 u_4}\,  u_1 \tens u_2  \ + \ b \,{du_4 \wedge
  du_3 \over u_3 u_4} \,  u_1 \tens u_2  \ + \ c \,{du_2 \wedge du_4  \over u_2 u_4}\,  u_1
  \tens u_3  \ + \ \mbox{  dihedral } \ =\ &0\ .\label{eq:6}
\end{align}
Here ``+ dihedral'' signifies the addition of all terms related by
dihedral transformations. Now we must consider the wedge terms. Since
$u_1= 1-u_3$, and $u_2 = 1-u_4$,   we have $du_1=-du_3$ and
$du_2=-du_4$. The minus sign disappears at the level of the symbol
(since it is blind to multiplication by constants) and
so there is  only one independent wedge product, $du_1
\wedge du_2$. For example we have:
\begin{align}
  du_1 \wedge du_3 = du_2 \wedge du_4 =0 \quad du_1 \wedge du_4 =
  du_1 \wedge du_2 \quad du_2 \wedge du_3 = - du_1 \wedge du_2 \quad
  \mbox{etc.}
\end{align}
So~(\ref{eq:6}a) becomes
\begin{align}\nonumber
   &(a-b)\,   d u_1 \wedge d u_2    \Big( {u_3 \tens u_4-u_4 \tens u_3 \over u_1 u_2}
   +  {u_1 \tens u_2-u_2 \tens u_1 \over u_3 u_4} \Big)
     \\
   +&(a-c)\,   d u_1 \wedge d u_2    \Big( {u_1 \tens u_4 -u_4 \tens u_1 \over u_2 u_3}
    +{u_3 \tens u_2 -u_2 \tens u_3 \over u_2 u_3} \Big)  \ =\ 0
\end{align}
and the other two equations are similar. The integrability constraint
therefore fixes $a=b=c$. This then yields the symbol of the function $\log
u_1 \log u_2\log u_3\log u_4$. So we conclude that the 2-loop 8-point
function~(\ref{eq:10}) is  fixed to be
\begin{align}
  \cR_8^{(2)} &= a \, (  \cR_{8;a}^{(2)}+ \cR_{8;b}^{(2)}+
  \cR_{8;c}^{(2)}) \ + \ 2 \cR_6^{(2)} \nonumber \\
&= a \ \log(u_1) \log(u_2) \log(u_3) \log(u_4)\ + \ 2 \cR_6^{(2)} \ ,
\end{align}
in agreement with the computed result~(\ref{r8}), with $a=-1/2$ and
$\cR_6^{(2)}=-\pi^4/36$. We could thus have derived the 2-loop 8-point result in
this case  with these reasonable assumptions, up to two unfixed
constants, one of which is
simply $\cR_6^{(2)}$.

\subsection{Lifting to $n$-point functions at two-loops}
\label{sec:lifting-n-point}

In the previous subsection we were able to derive the form of the
8-point 2-loop remainder function using some basic assumptions only
(dihedral symmetry, collinear limits, symbol made out of $u$'s). We now
wish to consider the lift to higher point functions. We find that the
result found in~\cite{2loop} is the unique function satisfying these
assumptions.
In~\cite{2loop} we assumed the result
consisted of logs of $u$ only, but now we can derive the same
result without this assumption.

Let us then analyse the most
general possible 10-point
function. This must be symmetric under dihedral symmetry and reduce to
the 8-point function under collinear limit. The most general solution
of this constraint is a ``particular solution'' together with the most
general dihedrally symmetric 10-point function which {\em vanishes} under
the collinear limit (the ``homogeneous solution''). So we have
\begin{align}
\cR_{10}^{(2)}&= \cR_{10;PS}+ \cR^{(2)}_{10;HS}\\
\cR_{10;PS}&=-\frac{1}{2} \Big(\log(u_1) \log(u_2) \log(u_3) \log(u_4) \,+\, {\rm cyclic}\Big)
\, -\, \frac{\pi^4}{12}\ ,
\end{align}
where we have $u_i:=u_{i i+4}$. Here $\cR_{10;PS}$ is a particular
solution of the collinear limit constraint. Indeed is is the
known 10-point
result from~\cite{2loop}. If we can show that $\cR^{(2)}_{10;HS}$
vanishes, then the solution is unique. Now $\cR^{(2)}_{10;HS}$ is a
dihedrally symmetric function which vanishes in any
collinear limit.
 In fact it is quite straightforward to see that no symbol exists
with these properties at 2 loops. All terms in the symbol of
$\cR^{(2)}_{10;HS}$
involves four $u$'s and so have the form $u_{i_1\,i_2} \tens
u_{i_3\,i_4} \tens u_{i_5\,i_6} \tens u_{i_7\,i_8} $. Now consider an
edge $j$ where $j \notin \{i_1, \dots i_8\}$ (clearly such an edge
exists at 10 or more points, but not at 8-points).  Now consider the
collinear limit occurring when $p_j\rightarrow 0$. This  implies that
$u_{j\,j+4}\rightarrow 1, \ u_{j\,j-4}\rightarrow 1$ and $u_{j-2\,j+2}
\rightarrow 0$ with all other $u$'s unconstrained (apart from via the
Y-system.) We can see that since $u_{j\,j+4}$ and $u_{j\,j-4}$ are not
in our tensor, it will not vanish in this collinear limit. Furthermore
there is no way for
different terms to combine to give vanishing contributions
either. We conclude that we can not obtain a collinear vanishing term
at 10 points.

So  the 2-loop 10 point function found in~\cite{2loop} is
the unique function whose symbol has cross-ratios as entries,
satisfying the correct collinear limits and dihedral symmetry.

The same analysis can be performed at all higher points and we thus
find that the solution~(\ref{rn}) found in~\cite{2loop} is the unique
$n$-point 2-loop
result satisfying our assumptions.

%%%%%%%%%%%%%%%%%%%%%%%%%%%%%%%%%%%%%%%%%%%%%%%%%%%%%%%%%%%%%%%%%%%%%%%%%%%%%%%%%%%%%%%%%%%
\section{The 3-loop octagon}
\label{sec:more-syst-appr}
%%%%%%%%%%%%%%%%%%%%%%%%%%%%%%%%%%%%%%%%%%%%%%%%%%%%%%%%%%%%%%%%%%%%%%%%%%%%%%%%%%%%%%%%%%%

In this section we describe our technique for applying constraints on
the form of the 8-point function at 3-loops in special kinematics.

In the following we will write the 8-point remainder as
\begin{align}
  \cR_8^{(3)}= F_8^{(3)} + 2 \cR_6^{(3)}
\end{align}
 where $\cR_6^{(3)}$ is constant in $1+1$ dimensions. Then according to the analysis in section~\ref{sec:collinear-limits} the function $F_8^{(3)}$ will vanish in the collinear limit.

All the 8-point remainder functions we have found can be written in the `sum of products' form
\begin{equation}
  \label{eq:1}
 F_8^{(3)}(u_1,u_2,u_3,u_4)=\, \sum_i \,{\rm const}_i[ f_i(u_1,u_3) g_i(u_2,u_4)+  g_i(u_1,u_3) f_i(u_2,u_4)]\, .
\end{equation}
Cyclic symmetry implies that $f$ and $g$ are symmetric functions
\begin{equation}
  \label{eq:2}
  f_i(u_1,u_3)=f_i(u_3,u_1) \qquad g_i(u_1,u_3)=g_i(u_3,u_1)
\end{equation}
and collinear limits imply that
\begin{equation}
  \label{eq:3}
  f_i(0,1)=0 \qquad g_i(0,1)=0\ .
\end{equation}

Since in the octagon case $u_3=1-u_1$ and $u_4=1-u_2$, the functions $f$ and $g$ are really functions of a single argument.
We thus will use a dual notation: when discussing symbols of $f$ and $g$, we will talk of $f(u,v)$ and $g(u,v)$, where
$u$ and $v$ are cross-ratios which satisfy $u+v=1$. On the other hand, when we reconstruct the actual functions we can choose to
 use the more appropriate single-argument definition:
\begin{equation}
  \label{eq:single-arg}
  f_i(u) \, :=\, f_i(u,1-u) \, , \qquad g_i(u)\,:=\, g_i(u,1-u) \,
\end{equation}
with (from~(\ref{eq:2}) $f_i(u)=f_i(1-u)$ and $g_i(u)=g_i(1-u)$, and (from the collinear limits~(\ref{eq:3})) $f_i(0)=0=f_i(1)$ and $g_i(0)=0=g_i(1)$.

The characteristic feature of the expression on the {\it r.h.s.} of \eqref{eq:1} is that
for each term in the sum the $u^+$ cross ratios, $u_1,u_3$ factorise from the $u^-$ cross ratios,
$u_2,u_4$.

To arrive at~\eqref{eq:1} we have started by writing down a general symbol
which by construction is a linear combination of weight-6 tensor products of the cross-ratios $u_1$, $u_2$, $u_3$ and $u_4$
(and not functions thereof as explained in section~\ref{sec:fud-assum}):
\begin{equation}
  \label{eq:1gen}
\cS[F_8^{(3)}(u_1,u_2,u_3,u_4)]=\,\sum_{i_1\ldots i_6} \,{\rm const}_{i_1\ldots i_6} \cdot u_{i_1} \tens u_{i_2}\tens u_{i_3} \tens u_{i_4}
\tens u_{i_5} \tens u_{i_6} \, .
\end{equation}
Next we imposed the requirement that the corresponding function should not explode (and in fact must vanish)
in any of the collinear limits, i.e. where
$u_i \rightarrow 0$ and $u_{i+2} \rightarrow 1$. This automatically requires that each tensor product must contain all four
cross-ratios $u_1$, $u_2$, $u_3$ and $u_4$. Indeed, to survive the collinear limit, $u_1 \rightarrow 0$ and $u_{3} \rightarrow 1$ for example,
whenever $u_1$ is present, there should also be a $u_3$ to regulate it, and the same applies for $u_2$ and $u_4$ for the limit
$u_2 \rightarrow 0$ and $u_{4} \rightarrow 1$ or vice versa. The second requirement is that the symbol in \eqref{eq:1gen} should respect
cyclic symmetry and parity, generated by~(\ref{eq:20})
which are the symmetries of the amplitude/Wilson loop. With these requirements the number of different constants
${\rm const}_{i_1\ldots i_6}$ in \eqref{eq:1gen} reduced to 195.
The final requirement we have imposed on \eqref{eq:1gen} is that it must be a symbol of a local function. This is known as the
$d^2 =0$ or integrability constraint, and described in section~\ref{sec:integr}. It implies that:
\begin{equation}
  \label{eq:dsq}
\sum_{i_1\ldots i_6} \,{\rm const}_{i_1\ldots i_6}  \, d\log(u_{i_k})\wedge d\log(u_{i_{k+1}})
\, u_{i_1} \ldots \tens u_{i_{k-1}}\tens u_{i_{k+2}} \ldots \tens u_{i_6} \,=\,0.
\end{equation}
for each $k$.
We found that implementing this constraint reduces the number of independent constants down to 13, and at the same time
imposes the `sum of products' functional form given by~\eqref{eq:1}.

We now come back to our starting point \eqref{eq:1} in order to describe the 13 functions explicitly.
The functions $f$ and $g$ must have transcendental weight 2 or more (since it must contain both $u_1$ and $u_3$ in its symbol in order not to vanish in the collinear limit)
and the product $f g$ sum must have weight 6. So we either have $g$
with weight 2 and $f$ with weight 4 or both $f,g$ have weight 3
each. Notice that in the case when $g$ has weight 2, the symmetry
in~(\ref{eq:2}) implies that the only possibility is $g(u,v)=\log(u) \log(v)$.

{\bf{Type a}}

This type has $g$ with weight 2 and $f(u,v)$ of weight 4,
consisting of 3 $u$'s and 1 $v$ in the symbol or vice versa.
 There are four
different possibilities, given by
\begin{eqnarray}
  g(u,v)&=& \log(u) \log(v) \nonumber\\
  \nonumber\\
   \cS[f(u,v)] &=& \left\{
    \begin{array}{l}
\cS[f_{a1}(u,v)] :=u \tens u \tens
  u \tens v  + v \tens v \tens
  v \tens u\\
\cS[f_{a2}(u,v)] :=u \tens u \tens
  v \tens u+v \tens v \tens
  u \tens v\\
\cS[f_{a3}(u,v)] :=u \tens v \tens
  u \tens u+v \tens u \tens
  v \tens v\\
\cS[f_{a4}(u,v)] :=v \tens u \tens
  u \tens u+u \tens v \tens
  v \tens v
\end{array}
\right.
\label{eq:4}
\end{eqnarray}

{\bf{Type b}}

This type has $g$ with weight 2 again and $f(u,v)$ of weight 4,
but this time consisting of 2 $u$'s and 2 $v's$ in the symbol.
 There are only three
different possibilities this time, given by
\begin{eqnarray}
  g(u,v)&=& \log(u) \log(v) \nonumber\\
  \nonumber\\
 \cS[f(u,v)] &=& \left\{
    \begin{array}{l}
\cS[f_{b1}(u,v)] :=u \tens u \tens
  v \tens v  + v \tens v \tens
  u \tens u\\
\cS[f_{b2}(u,v)]:=u \tens v \tens
  u \tens v+v \tens u \tens
  v \tens u\\
\cS[f_{b3}(u,v)]:=u \tens v \tens
  v \tens u+v \tens u \tens
  u \tens v
\end{array}
\right.
\label{eq:4b}
\end{eqnarray}

{\bf{Type c}}

Finally we have type c functions in which both  $f$ and $g$ have
weight 3.
 There are three  possibilities for both $f$ and $g$, given by
\begin{equation}
 \cS[ f(u,v)] \mbox{ or } \cS[g(u,v)] = \left\{
    \begin{array}{l}
 \cS[ f_{c1} (u,v)] := u \tens u \tens
  v   + v \tens v \tens u\\
 \cS[ f_{c2} (u,v)] :=u \tens v \tens
  u +v \tens u \tens  v \\
 \cS[ f_{c3} (u,v)] :=u \tens v \tens v+v \tens u \tens u
\end{array}
\right.
\label{eq:4c}
\end{equation}
yielding 6 possible functions
\begin{eqnarray}
  \nonumber
&&  f_{c1}(u_1,u_3)f_{c1}(u_2,u_4)\,;\\\nonumber
 && f_{c1}(u_1,u_3)f_{c2}(u_2,u_4)+f_{c2}(u_1,u_3)f_{c1}(u_2,u_4)\,;\\\nonumber
  &&f_{c1}(u_1,u_3)f_{c3}(u_2,u_4)+f_{c3}(u_1,u_3)f_{c1}(u_2,u_4)\,;\\\nonumber
  &&f_{c2}(u_1,u_3)f_{c2}(u_2,u_4)\,;\\\nonumber
  &&f_{c2}(u_1,u_3)f_{c3}(u_2,u_4)+f_{c3}(u_1,u_3)f_{c2}(u_2,u_4)\,;\\
  &&f_{c3}(u_1,u_3)f_{c3}(u_2,u_4)\, .
\label{eq:5}
\end{eqnarray}

At this point we have in total 13 combinations: 4 from type-a, 3 from type-b and 6 from type-c above.
We have already imposed the dihedral symmetry on the answer and have built into it the requirement that our 8-point expression
must vanish in the collinear limit as required in the special kinematics and these are of course the 13 functions we found using the computer based method described around~(\ref{eq:dsq}).

However, we have not yet checked that
all the functions we have constructed so far vanish {\it sufficiently slowly} in the collinear limit. We will show now
that three of our structures, $f_{a1}$, $f_{a2}$ and $f_{b_1}$,
are actually more singular in the collinear limit than allowed, and will have to be discarded, reducing the number of
allowed combinations by 3.
In fact, using the near-collinear OPE, the authors~\cite{gmsv} have deduced the leading behaviour of the
three-loop result,
\begin{equation}
\lim_{u_1 \rightarrow 0}
F_8^{(3)}(u_1,u_2,u_3,u_4) =\, \log^2(u_1) \log(u_3) \cdot F_3 (u_2,u_4) + O(\log(u_1))\ .
\label{eq:7a}
\end{equation}
where $F_3 (u_2,u_4)$ is known and was written in~\cite{gmsv} in the form
\begin{eqnarray}
F_3 (u_2,u_4) &=& -2\Li_3(1 - 1/u_4) + \log(u_2/u_4) \Li_2(1 - 1/u_4) +
 \frac{4}{3} \log^3(u_4) \nonumber\\
 &+& 2 \log(u_2/u_4) \log^2(u_4) +
 \frac{1}{2} \log^2(u_2/u_4) \log(u_4) + \frac{\pi^2}{6} \log(u_4)\, .
 \label{eq:7b}
\end{eqnarray}
We will return to the function $F_3(u_2,u_4)$ below, but first we concentrate on the $u_1$-, $u_3$-dependence in \eqref{eq:7a}.
This equation implies that the answer $\propto \log^2(u_1) \log(u_3)$
in the limit $u_1 \rightarrow 0$, $u_3:=1-u_1 \rightarrow 1$. This functional form rules out $f_{a1}$ and $f_{a2}$ since
\begin{eqnarray}
&&\lim_{u_1\rightarrow 0} \cS^{-1}\big(u_1\tens u_1 \tens u_1 \tens u_3\big) \  \sim \ \log^3(u_1) \log(u_3) \, ,\nonumber\\
&&\lim_{u_1\rightarrow 0} \cS^{-1}\big(u_1\tens u_1 \tens u_3 \tens u_1 \big)\  \sim \ -
\log^2(u_1) \Li_2(u_1)\, , \nonumber
\end{eqnarray}
giving functions with the wrong asymptotic properties.
The function $f_{b_1}$ is ruled out for the same reason.

Explicit expressions for the remaining seven functions
\begin{equation}
f_{ai},f_{bi},f_{ci}(u)\,:=\, f_{ai},f_{bi},f_{ci}(u,v\equiv 1-u )
\end{equation}
 can now be
straightforwardly reconstructed from their symbols \eqref{eq:4}-\eqref{eq:4c} by taking into account the constraint on the variables $v=1-u$,
and the properties of the symbol. We find
\begin{eqnarray}
\nonumber
  f_{a3}(u,v)&=& 3\Li_4(u) -\Li_3(u)\log(u)+ 3\Li_4(v) -\Li_3(v)\log(v) -\frac{\pi^4}{30}\, ,
\\\nonumber
f_{a4}(u,v)&=& -\Li_4(u) -\Li_4(v)+\frac{\pi^4}{90}\, ,
\\\nonumber
f_{b2}(u,v)&=& \left(\Li_3(u)-\zeta_3\right) \log(v)-\Li_2(u)\Li_2(v) +\log^2(u)\log^2(v)
+\left(\Li_3(v)-\zeta_3\right) \log(u)\, ,
\\\nonumber
f_{b3}(u,v)&=& -\left(\Li_3(u)-\zeta_3\right) \log(v)+\Li_2(u)\Li_2(v) -\frac{1}{2}\log^2(u)\log^2(v)
-\left(\Li_3(v)-\zeta_3\right) \log(u)\, ,
\\\nonumber
f_{c1}(u,v)&=& -\Li_3(u) -\left(\Li_2(v)-\frac{\pi^2}{6}\right) \log(u)-\frac{1}{2}\log(v)\log^2(u)
\\\nonumber
&&-\Li_3(v)-\left(\Li_2(u)-\frac{\pi^2}{6}\right) \log(v)- \frac{1}{2}\log(u)\log^2(v) +\zeta_3\, , \\\nonumber
f_{c2}(u,v)&=& 2\Li_3(u) +\left(\Li_2(v)-\frac{\pi^2}{6}\right) \log(u)+\log(v)\log^2(u)
\\\nonumber
&&+2\Li_3(v)+\left(\Li_2(u)-\frac{\pi^2}{6}\right) \log(v)+\log(u)\log^2(v) -2\zeta_3\, , \\
f_{c3}(u,v)&=& -\Li_3(v) -\Li_3(u)+\zeta_3\, ,
\label{eq:funs}
\end{eqnarray}
where the constants on the {\it r.h.s} are determined from the requirement that all functions must vanish in the collinear limit.

We can now further constrain 3 more coefficients of our general expression by making use of the function $F_3(u_2,u_4)$. First we find another equivalent form for the function in \eqref{eq:7b} so that its arguments on the {\it r.h.s.}
are just the cross-ratios $u_2$ and $u_4$:
\begin{align}
F_3 (u_2,u_4) =& 2 \Li_3(u_2) + \left(\Li_2(u_4) - \frac{\pi^2}{6}\right) \log(u_2)+\frac{3}{2} \log(u_4) \log^2(u_2) \nonumber\\
 +& 2 \Li_3(u_4) + \left(\Li_2(u_2) - \frac{\pi^2}{6}\right) \log(u_4)+\frac{3}{2} \log(u_2) \log^2(u_4)- 2\zeta_3
 \label{eq:7c}
\end{align}
This function has the same symbol as \eqref{eq:7b} (note that as always $u_2+u_4=1$ at 8-points) and moreover
we checked that the two functions agree numerically. Now we notice that $F_3 (u_2,u_4)$ in \eqref{eq:7c} is
just a linear combination of our functions $f_{c1}$, $f_{c2}$ and $f_{c3}$.
In other words, the GMSV condition takes the form
\begin{equation}\label{eq:7}
\lim_{u_1 \rightarrow 0}
F_8^{(3)}(u_1,u_2,u_3,u_4)=\log^2(u_1) \log(u_3)  \Big[
f_{c1}(u_2,u_4)+2f_{c2}(u_2,u_4)+f_{c3}(u_2,u_4)
 \Big] + O(\log(u_1))\ .
\end{equation}
We conclude that
the coefficients in front of 3 of the 6 $c$-type functions listed
in~(\ref{eq:5}) are fixed.
We note that the fact that the {\it r.h.s.} of Eq.~\eqref{eq:7c} can be presented entirely in terms of simple functions of
of cross-ratios $(u_2,u_4)$, and more specifically that the symbol of $F_3 (u_2,u_4)$ is the tensor product of $u$ variables,
gives a self-consistency check on our fundamental assumption that the symbol of the full answer is made out of $u$'s.

We can now write the most general function consistent
with all available conditions:
\begin{align}
  \label{eq:8}
 & F_8^{(3)}(u_1,u_2,u_3,u_4)\nonumber\\
&= \log u_1 \log u_3 \Big[ \alpha_1 \,f_{a3}(u_2,u_4) +\alpha_2
  \,f_{a4}(u_2,u_4) +\alpha_3 \,f_{b2}(u_2,u_4)  +\alpha_4
  \,f_{b3}(u_2,u_4) \Big] \nonumber \\
&\ +\alpha_5  f_{c2}(u_1,u_3)f_{c2}(u_2,u_4)+
\alpha_6
f_{c2}(u_1,u_3)f_{c3}(u_2,u_4)+
 \alpha_7  f_{c3}(u_1,u_3)f_{c3}(u_2,u_4) \nonumber\\
&\ + f_{c1}(u_1,u_3)\Big[ \half f_{c1}(u_2,u_4)+
2 f_{c2}(u_2,u_4)+f_{c3}(u_2,u_4)\Big]\nonumber\\
& \ + ( u_1 \leftrightarrow u_2, u_3 \leftrightarrow u_4)
\end{align}
Thus we have obtained an analytic expression for the 3-loop contribution to the 8-point amplitude which contains 7 free
constants $\alpha_i$. It is remarkable that the 3-loop octagon in special 2d kinematics can be written in such a compact form
and involving only classical polylogarithms of degree $\le 4$ and logarithms. It is clearly important to further constrain
at least some of the yet undetermined 7 coefficients in the expression above. It would be interesting to investigate whether
one can fix some of the $\alpha$'s by going to the BFKL limit of the 8-point amplitude in the special kinematics -- for the lower hexagon case,
this procedure has
reduced the number of free constants at 3-loops in general kinematics, as was shown very recently in \cite{Dixon:2011pw}. We have not attempted
to generalise their approach to the octagon case considered here.

In the
following section we will outline the procedure of finding the general uplift to 10-points. This approach is general and
conceptually there are no restrictions for continuing to an arbitrary high number of 2n-points.

Finally, the fact that not just the symbol, but the functional form of the 3-loop 8-point result is now known,
one would be able to determine the coefficients and check the validity of the above approach against numerical results
at a few fixed values of the cross-ratios, whenever these results become available.

\section{Lifting the 3-loop octagon to higher polygons}
\label{sec:lift-high-points}

\subsection{Constructing the decagon: part 1}

We will now show how to uplift the 8-point function to 10 points guided by the collinear limits.
Here we will construct a `particular solution' for the 10-point polygon remainder, which is just consistent
with the collinear limits. In the following subsection we will obtain the general solution by determining
all 10-point structures which vanish in the collinear limit.
Similarly to 8-points, we will write 
\begin{align}
  \cR_{10}^{(3)}= F_{10}^{(3)} + 3 \cR_{6}^{(3)}
\end{align}
 so that under the collinear limits described in section~\ref{sec:collinear-limits} we have simply $F_{10}^{(3)} \ \rightarrow \ F_{8}^{(3)}$. We will then consider the various contributions to $F_{10}^{(3)}$.

To begin with we consider the 8-point type-a and type-b functions. As explained in Section~\ref{sec:more-syst-appr}, they are of the form,
\begin{equation}
  \label{eq:9}
F_{ab\,8}=\,  \log (u_1) \log (u_3)\, f_{ab}(u_2,u_4) \,+\, \log (u_2) \log (u_4) \, f_{ab}(u_1,u_3)\ ,
\end{equation}
where
\begin{equation}
\label{eq:gab}
f_{ab}(u_2,u_4) :=\, \alpha_1 \,f_{a3}(u_2,u_4) +\alpha_2
  \,f_{a4}(u_2,u_4) +\alpha_3 \,f_{b2}(u_2,u_4)  +\alpha_4
  \,f_{b3}(u_2,u_4)\, ,
\end{equation}
as can be seen from the first line on the {\it r.h.s.} of \eqref{eq:8}.
We now find the lift of this expression to 10-points, it  turns out that this is quite
straightforward.

The 10-point function is  supposed to reduce under the collinear limit $u_5 \rightarrow 1$, $u_7\rightarrow
0, u_9 \rightarrow 1$ to the corresponding 8-point function with $u_4$ replaced by $u_4 u_{10}$.
So in this case we are supposed to  get
\begin{equation}
\log (u_1) \log (u_3)\, f_{ab}(u_2,u_4 u_{10})\,+\, \log (u_2) \log (u_4 u_{10})\, f_{ab}(u_1,u_3)
\label{eq:12}
 \end{equation}
in the collinear limit.

To achieve an uplift consider the function
\begin{equation}
\label{eq:10pt-ab-sp-1}
  \log (u_1) \log (u_3) \ f_{ab}(u_2,u_4 u_{10})  + \mbox{cyclic }\ .
 \end{equation}
One can easily check, using~(\ref{eq:2}) and~(\ref{eq:3}) that this function
reduces under the collinear limit correctly to \eqref{eq:12}.
Indeed, the three terms, corresponding to $i=1$, $i=2$ and $i=10$ of
\begin{equation}
\label{eq:10pt-ab-sp}
 F_{ab\,10}=\,\sum_{i=1}^{10} \log (u_i) \log (u_{i+2}) \ f_{ab}(u_{i+1},u_{i+3} u_{i-1})\, ,
 \end{equation}
combine to the two terms in \eqref{eq:12}. All the remaining terms in the sum in \eqref{eq:10pt-ab-sp} vanish
in this limit.

Note that in the octagon case, the functions $f_{ab}(u_2,u_4)$ in \eqref{eq:gab} were in fact functions of a {\it single} variable
$u_2$, since for the octagon $u_4=1-u_2$ (and $u_3=1-u_1$). Hence it would have been more appropriate to define
$f_{ab}(u_2):=f_{ab}(u_2,1-u_2)$. The question arises as what is the meaning of the function $f_{ab}(u_2,u_4 u_{10})$ appearing
in the decagon case in \eqref{eq:10pt-ab-sp-1} and \eqref{eq:10pt-ab-sp}? In fact it is the same function of a single variable $u_2$
just as for $n=8$.
The Y-system for the decagon \eqref{eq:10Ysoln}
allows us to rewrite the products as $u_4 u_{10}=1-u_2$, and since
\begin{equation}
\label{eq:gprod}
f_{ab}(u_2,u_4 u_{10}) \,=\, f_{ab}(u_2,1-u_2) \,:=\, f_{ab}(u_2) \, .
\end{equation}
Furthermore, since the symbol for each $f_{ab}(u)$
in~\eqref{eq:4}-\eqref{eq:4c} involves only $u$
and $1-u$ (which can be rewritten as  a product of $u$'s) then making
use of the product rule for symbols~(\ref{eq:19}), we see that the symbols of these functions are indeed made out of tensor products of the $u$'s alone, with no functions of $u$ appearing,
 as required by our fundamental assumption in special kinematics. We
 can write
\begin{equation}
\label{eq:10pt-ab-sp-s}
 F_{ab\,10}=\,  \log (u_1) \log (u_3) \ f_{ab}(u_2)  + \mbox{cyclic }\ .
 \end{equation}
So in this way we can immediately uplift the type-a and type-b 8-point
functions to 10 points. To find the general solution to the 10-point structure one would need to add to
Eq.~\eqref{eq:10pt-ab-sp-s} (and to Eq.~\eqref{eq:10pt-c-sp} below)
also the general set of functions which vanish in the collinear limit
(this is a new possibility at 3 loops which couldn't occur at 2 loops
as detailed in section~\ref{sec:lifting-n-point}).  This
will be done in the following subsection.

Now consider the type-c 8-point functions. These are of the form
\begin{equation}
\label{eq:8-c}
F_{c\,8}=\, \sum \, f_c(u_1,u_3)g_c(u_2,u_4) + g_c(u_1,u_3)f_c(u_2,u_4)  \, ,
 \end{equation}
with $f$ and $g$ each of
weight-3. Equation \eqref{eq:8-c} corresponds to the second-through-last lines on the {\it r.h.s.} of \eqref{eq:8}.

The corresponding 10-point function we are trying to obtain
is therefore supposed to reduce under the collinear limit, $u_5 \rightarrow 1$, $u_7 \rightarrow 0$, $u_9 \rightarrow 1$
to
$f_c(u_1,u_3)g_c(u_2,u_4u_{10}) + g_c(u_1,u_3)f_c(u_2,u_4u_{10})$.

Again this is fairly straightforward to achieve, we simply take
\begin{align}
F_{c \,10}=\,& \frac{1}{2}\,f_c(u_1)\Big(g_c(u_2)-g_c(u_4)+g_c(u_6)-g_c(u_8)+g_c(u_{10})\Big)
+ \mbox{cyclic }\ .
\label{eq:10pt-c-sp}
\end{align}
where we have once again defined the single-argument functions,
\begin{equation}
f_{c}(u):=f_{c}(u,1-u)\, , \qquad g_{c}(u):=g_{c}(u,1-u)\, ,
\end{equation}
so that $f_c(u)=f_c(1-u)$ and $g_c(u)=g_c(1-u).$

To see that the {\it r.h.s} of \eqref{eq:10pt-c-sp} reduces to the desired expression in the collinear limit,
we note that when we take $u_5 \rightarrow 1$, $u_7 \rightarrow 0$, $u_9 \rightarrow 1$ we also have automatically
$u_3=1-u_1$, as must be the case for the octagon. Thus in the
collinear limit, $f_c(u_1)=f_c(u_3)$ so that the first and the third terms in
the cyclic permutation have $g_c(u_4)$, $g_c(u_6)$, $g_c(u_8)$ and
$g_c(u_{10})$ cancelled and amount to $f_c(u_1) g_c(u_2)$, while the
second term produces the other required factor, $f_c(u_2) g_c(u_1)$. The remaining cyclic permutations
in \eqref{eq:10pt-c-sp} vanish in the limit.
Furthermore, as before, the functions $f_c(u)$ and $g_c(u)$ are  made out of tensor products of $u$'s alone.

\subsection{Constructing the decagon: part 2. Collinear-vanishing 10-point functions}
\label{sec:coll-vanish-10}

We have uplifted the 8-point function to 10 points, but to what extent
is this unique? There exist collinear-vanishing 3-loop functions at
10-points and these can never be detected by this uplift. The most
general possible 10-point function is the function uplifted from 8
points plus the most general collinear vanishing 10-point function.

We approach the problem of finding the most general collinear
vanishing 10-point function in two independent ways. Firstly we work systematically: using
a computer, we write down the most general cyclic and parity
symmetric, collinear vanishing symbol made of tensor products of $u$'s. Then we impose the
integrability constraint \eqref{eq:integr}. This gives 888 constraints thus leaving just
12 collinear vanishing functions.

The second method starts with the
assumption that the collinear vanishing function has the form
\begin{equation}
  \label{eq:14}
  f(u_i^+)g(u_i^-) \ + \ \mbox{cyclic} \ + \ \mbox{parity}\ .
\end{equation}
Now we analyse the possible functions $f,g$. These functions must
themselves vanish in any collinear limit. To do this they must have
weight 3 or more and each term must contain 3 consecutive $u_i^\pm$ eg $u_1^+,
u_2^+,u_3^+= u_1,u_3,u_5$.
So since the same conditions are  true for both  functions $f,g$,
and the total weight is 6, they must both have weight 3. Now writing out the most general such
symbol for $f$ (or equivalently $g$) and imposing the integrability
constraint we 
find there are just 11 possibilities which come in 3 types. These are
not too hard to find analytically:
\begin{align}
  \label{eq:15}
 f_1(u^\pm_1,u^\pm_2,u^\pm_3) &= \log(u^\pm_1) \log(u^\pm_2)
 \log(u^\pm_3) \nonumber\\
  f_2(u^\pm_1,u^\pm_2,u^\pm_3)&= \log(u^\pm_2)\Big(\Li_2(u^\pm_1)
  -\Li_2(1-u^\pm_2)+\Li_2(u^\pm_3)  -\pi^2/6  \Big) \nonumber\\
f_3(u_i^\pm)&= \sum_{i=1}^5 \Big(\Li_3(u_i^\pm)-\Li_3(1-u_i^\pm)\Big)-\zeta_3\ .
\end{align}
Here $f_1$ and $f_2$  give 5 independent functions via cyclic permutations of the
arguments, whereas $f_3$ is cyclically symmetric giving only 1
independent function, thus we have 11
functions in total.
We can now combine these together to obtain a total of 12 independent  weight 6 collinear
vanishing 10 point function as follows:
\begin{align}
  f_1(u_1,u_3,u_5)f_1(u_2,u_4,u_6) \ + \ \mbox{cyclic} \ + \ \mbox{parity}\ \nonumber \\
   f_1(u_1,u_3,u_5)f_1(u_4,u_6,u_8)\ + \ \mbox{cyclic} \ + \ \mbox{parity}\ \nonumber \\
   f_1(u_1,u_3,u_5)f_1(u_6,u_8,u_{10})\ + \ \mbox{cyclic} \ + \ \mbox{parity}\ \nonumber \\
   f_1(u_1,u_3,u_5)f_2(u_2,u_4,u_6)\ + \ \mbox{cyclic} \ + \ \mbox{parity}\ \nonumber \\
   f_1(u_1,u_3,u_5)f_2(u_4,u_6,u_8)\ + \ \mbox{cyclic} \ + \ \mbox{parity}\ \nonumber \\
   f_1(u_1,u_3,u_5)f_2(u_6,u_8,u_{10})\ + \ \mbox{cyclic} \ + \ \mbox{parity}\ \nonumber \\
   f_2(u_1,u_3,u_5)f_2(u_2,u_4,u_6)\ + \ \mbox{cyclic} \ + \ \mbox{parity}\ \nonumber \\
   f_2(u_1,u_3,u_5)f_2(u_4,u_6,u_8)\ + \ \mbox{cyclic} \ + \ \mbox{parity}\ \nonumber \\
   f_2(u_1,u_3,u_5)f_2(u_6,u_8,u_{10})\ + \ \mbox{cyclic} \ + \ \mbox{parity}\ \nonumber \\
  f_1(u_1,u_3,u_5)f_3(u_i^-) \ + \ \mbox{cyclic} \ + \ \mbox{parity}\ \nonumber \\
   f_2(u_1,u_3,u_5)f_3(u_i^-)\ + \ \mbox{cyclic} \ + \ \mbox{parity}\ \nonumber \\
   f_3(u_1,u_3,u_5)f_3(u_i^-)\ + \ \mbox{cyclic} \ + \ \mbox{parity}\
   \label{eq:last}
\end{align}
Our expression for the decagon remainder function is obtained by adding together
equations \eqref{eq:10pt-ab-sp-s}, \eqref{eq:10pt-c-sp} (with constants $\alpha_1,\ldots,\alpha_7$)
and 12 contributions from \eqref{eq:last}. We have thus constructed the general analytic expression for the
decagon which contains 19 as yet undetermined constant coefficients.

One obvious question is if the current understanding of the
near-collinear OPE at 10-points could restrict the
function further. Unfortunately this is not  the case.  None of the
collinear vanishing terms found here contribute to the OPE
(at the order at which this is currently understood) and nor does the
function $F_{ab\ 10}$. Indeed only
those functions in $F_{c\ 10}$
whose coefficients have already been fixed by the 8-point OPE are
detectable by the 10-point OPE, thus providing a consistency check, but no
new information.

\section{Conclusions}

The main results of this paper are derived from a single fundamental assumption of what are the correct variables
of the Wilson loop symbol in the special kinematics. We have postulated that these variables are given by the conformal cross-ratios
$u_{ij}$ so that the symbol is a sum of tensor products of $u_{ij}$.

Based on this constraint on the symbol, and using the symmetries of the system together with collinear limits, we have re-derived the
2-loop $n$-point analytic expressions for general (even) $n$ in agreement with the previously known results of~\cite{dds8,2loop}.
Our purpose was to achieve this without performing the direct perturbative computation (which was carried out in
\cite{dds8} for $n=8$), whilst making a weaker assumption than was made in \cite{2loop} that only $\log(u)$ can appear in the two-loop answer.

We then applied this strategy at 3-loops in section~\ref{sec:more-syst-appr} where we have determined the functional form
of the $8$-point Wilson loop answer. Our analytic result has a very compact form and is expressed in terms of logarithms and
classical polylogarithms of cross-ratios only. After imposing the constraint arising from the near-collinear OPE of~\cite{gmsv}
we ended up with 7 so far undetermined constant coefficients $\alpha_1,\ldots,\alpha_7$. Our final result for the octagon at 3-loops
is given by
\begin{align}
  \label{eq:8-new}
  F_8^{(3)}=& \log u_1 \log (1-u_1) \Big[ \alpha_1 \,f_{a3}(u_2) +\alpha_2
  \,f_{a4}(u_2) +\alpha_3 \,f_{b2}(u_2)  +\alpha_4
  \,f_{b3}(u_2) \Big] \nonumber \\
&+\alpha_5  f_{c2}(u_1)f_{c2}(u_2)+
\alpha_6
f_{c2}(u_1)f_{c3}(u_2)+
 \alpha_7  f_{c3}(u_1)f_{c3}(u_2) \nonumber\\
&+ f_{c1}(u_1)\Big[ \frac{1}{2} f_{c1}(u_2)+
2 f_{c2}(u_2)+f_{c3}(u_2)\Big]\nonumber\\
& + ( u_1 \leftrightarrow u_2)
\end{align}
with the $f_{a}$, $f_b$ and $f_c$ functions defined in \eqref{eq:funs}.

Our strategy also works for higher polygons. Following the uplift of the 8-point answer, we have constructed the
general analytic expression for the 3-loop decagon.
Our result for the 3-loop decagon is
given by
\begin{eqnarray}
\label{eq:decagon}
 F_{10}^{(3)}&=&  \sum_{k=1}^{12}\, \beta_k \, \phi_k \, +\,
 \log (u_1) \log (u_3) \ \tilde f_{ab}(u_2) \\
&+&\frac{1}{2}\sum_{i=1}^3 \,f_{c_i}(u_1)\Big(\tilde f_{c_i}(u_2)-\tilde f_{c_i}(u_4)+\tilde f_{c_i}(u_6)-\tilde f_{c_i}(u_8)+\tilde f_{c_i}(u_{10})\Big)
+ \mbox{cyclic } \nonumber
 \end{eqnarray}
where
$\phi_k$ are the 12 combinations on the {\it r.h.s.} of \eqref{eq:last} and
\begin{eqnarray}
\tilde f_{ab}(u) &=& \alpha_1 \,f_{a3}(u) +\alpha_2 \,f_{a4}(u) +\alpha_3 \,f_{b2}(u)  +\alpha_4 \,f_{b3}(u)\, , \nonumber\\
\tilde f_{c_1}(u) &=& \half \,f_{c_1}(u) + 2\,f_{c_2}(u) + \,f_{c_3}(u) \, , \nonumber\\
\tilde f_{c_2}(u) &=& \alpha_5 \,f_{c_2}(u) + \alpha_6 \,f_{c_3}(u) \, , \nonumber\\
\tilde f_{c_3}(u) &=& \alpha_7 \,f_{c_3}(u) \, ,
\label{eq:gab-new}
\end{eqnarray}
with the $f_{a}$, $f_b$ and $f_c$ functions collected in \eqref{eq:funs}. The 19 free constants are
$\beta_1,\ldots,\beta_{12}$ and $\alpha_1,\ldots,\alpha_7$.

One should also bear in mind that one can always add to any 3-loop
remainder $\pi^2$ times the 2-loop remainder (which  as has been seen
is uniquely fixed by our considerations). Such a possibility can never
be ruled out from our considerations, since we know that this
satisfies all the requirements we are insisting upon.
so in other words we can always add
\begin{equation}
  \label{eq:13}
\cR_n^{(3)} \ \rightarrow \ \cR_n^{(3)} \ + \ k \, \pi^2 \cR_n^{(2)}\ .
\end{equation}
But this is the only possible lower transcendental function we can add.

In principle, there are no obstacles in continuing to uplift these
3-loop results to higher points. An important point here  is that
there are no collinear vanishing functions beyond 12 points within our
ansatz (and more
generally at $l$ loops beyond $4l$ points.) This can be seen easily
from the point of view of the symbol, there are too many edges
(compared with the rank of the symbol-tensor) for one to ensure that each term
in the tensor always contains a cross-ratio approaching unity in the
limit,  which is the only way to 
kill this term. We conclude that once the 12 point 3-loop remainder is
known, and more generally the $4l$ point $l$-loop remainder, the
uplift to all higher points is unique.

It will be also interesting to
continue this programme to higher loops.

\bigskip
\bigskip

{\bf Acknowledgements}

We would like to thank Herbert Gangl and Claude Duhr for stimulating
discussions. We would also like to thank Timothy Goddard for a careful reading of the manuscript.
 VVK would like to thank the organisers and participants of the Amplitudes programme at KITP
and the hospitality of the Aspen Center for Physics.

\end{document}